\documentclass[a4paper]{article}
\usepackage{fullpage,xspace,verbatim,graphicx}
\usepackage{amsmath,amssymb}
\usepackage{color,cite,authblk}
\usepackage[squaren]{SIunits}
\usepackage[hidelinks]{hyperref}

\newcommand{\be}{\[}
\newcommand{\ee}{\]} 
\newcommand{\dif}{\mathrm{d}}
\newcommand{\me}[1]{\mathrm{e}^{#1}}

\addunit{\tcid}{TCID_{50}}
\addunit{\pfu}{PFU}
\addunit{\idose}{infectious\ doses}
\addunit{\idnew}{SIN}

\newcommand{\ID}{ED\xspace}
\newcommand{\SKlong}{Spearman-K{\"a}rber\xspace}
\newcommand{\RMlong}{Reed-Muench\xspace}
\newcommand{\SK}{SK\xspace}
\newcommand{\RM}{RM\xspace}
\newcommand{\idfif}{\ensuremath{\text{ID}_{50}}\xspace}
\newcommand{\progurl}{\url{https://midsin.physics.ryerson.ca}\xspace}
\newcommand{\giturl}{\url{https://github.com/cbeauc/midSIN}\xspace}
\newcommand{\newprog}{\textbf{midSIN}\xspace} 
\newcommand{\Cvir}{\ensuremath{C_\mathrm{inf}}\xspace}
\newcommand{\lCvir}{\ensuremath{\ell_\mathrm{Cinf}}\xspace}
\newcommand{\lidnew}{\ensuremath{\log_{10}(\idnew/\milli\liter)}\xspace}
\newcommand{\lten}{\ensuremath{\log_{10}}\xspace}

\newcommand{\Vinoc}{\ensuremath{V_\text{inoc}}\xspace}
\newcommand{\Vtot}{\ensuremath{V_\text{sample}}\xspace}
\newcommand{\Vvir}{\ensuremath{V_\text{vir}}\xspace}
\newcommand{\Nvir}{\ensuremath{N_\text{vir}}\xspace}

\newcommand{\qnoinf}{\ensuremath{q_\text{noinf}}\xspace}
\newcommand{\Dil}{\ensuremath{\mathcal{D}}\xspace}

\newcommand{\Pcal}{\ensuremath{\mathcal{P}}\xspace}
\newcommand{\Qcal}{\ensuremath{\mathcal{Q}}\xspace}
\newcommand{\qi}{\ensuremath{q_i}\xspace}
\newcommand{\Ppri}{\ensuremath{\Pcal_\text{prior}}\xspace}
\newcommand{\Upost}{\ensuremath{\mathcal{U}_\text{post}}\xspace}

\title{Time to revisit the endpoint dilution assay and to replace \tcid\ and \pfu\ as measures of a virus sample's infection concentration}

\author[1]{Daniel Cresta}
\author[2]{Donald C.\ Warren}
\author[1]{Christian Quirouette}
\author[3]{Amanda P.\ Smith}
\author[3]{Lindey C.\ Lane}
\author[3]{Amber M.\ Smith}
\author[1,2]{Catherine A.\ A.\ Beauchemin}

\affil[1]{Department of Physics, Ryerson University, Toronto, ON, M5B 2K3, Canada}
\affil[2]{Interdisciplinary Theoretical and Mathematical Sciences (iTHEMS) program, RIKEN, Wako-shi, Saitama, 351-0198, Japan}
\affil[3]{Department of Pediatrics, University of Tennessee Health Science Center, Memphis, TN, 38163, USA}

\date{\today}

\begin{document}


\maketitle

\begin{abstract}
The infectivity of a virus sample is measured by the infections it causes, via a plaque or focus forming assay (\pfu\ or FFU) or an endpoint dilution (\ID) assay (\tcid, CCID$_{50}$, EID$_{50}$, etc., hereafter collectively \idfif). The counting of plaques or foci at a given dilution intuitively and directly provides the concentration of infectious doses in the undiluted sample. However, it has many technical and experimental limitations. For example, it is subjective as it relies on one's judgement in distinguishing between two merged plaques and a larger one, or between small plaques and staining artifacts. In this regard, \ID assays are more robust because one need only determine whether or not infection occurred. The output of the \ID assay, the 50\% infectious dose (\idfif), is calculated using either the \SKlong (1908\textbar1931) or \RMlong (1938) mathematical approximations. However, these are often miscalculated and their approximation of the \idfif cannot be reliably related to the infectious dose. Herein, we propose that the plaque and focus forming assays be abandoned, and that the measured output of the \ID assay, the \idfif, be replaced by a more useful measure we coined \emph{specific infections} (\idnew). We introduce a free, open-source web-application, \newprog, that computes the \idnew\ concentration in a virus sample from a standard \ID assay, requiring no changes to current experimental protocols. We use \newprog to analyze sets of influenza and respiratory syncytial virus samples, and demonstrate that the \idnew/mL of a sample reliably corresponds to the number of infections a sample will cause per unit volume. The \idnew/mL concentration of a virus sample estimated by \newprog, unlike the \idfif/mL, can be used directly to achieve the desired multiplicity of infection. Estimates obtained with \newprog are shown to be more accurate and robust than those obtained using the \RMlong and \SKlong approximations. The impact of \ID plate design choices (dilution factor, replicates per dilution) on measurement accuracy is also explored. The simplicity of \idnew\ as a measure and the greater accuracy provided by \newprog make them an easy and superior replacement for the \pfu, FFU, \tcid\ and other ID$_{50}$ measures. We hope to see their universal adoption to measure the infectivity of virus samples.
\end{abstract}

\section{Introduction}

The progression of a virus infection \emph{in vivo} or \emph{in vitro}, or the effectiveness of therapeutic interventions in reducing viral loads, are monitored over time through sample collections to measure changes (increases or decreases) in virus concentrations. As such, accurate measurement of the concentration in a sample is critical to study and manage virus infections. The most direct method is to count individual virions as observed under an electron microscope. However, this technique is costly, time consuming, and largely destructive of the samples, and is thus almost never used. Viral RNA can be counted via quantitative polymerase chain reaction (qPCR), a method that amplifies a specific virus genome segment (RNA or DNA) within the sample over multiple cycles. The growth curve resulting from successive amplification cycles, compared against the standard curve for a sample of known concentration, provides an estimate of the number of viral segments in the sample. The major limitation of this method is that it measures not only viral RNA from intact virions, only some of which are infection-competent, but also debris from apoptotic or lysed cells, and antibody- or antiviral-neutralized virions, which misrepresents the effective virion concentration. For this reason, a count of infectious particles rather than, or in addition to, total viral genome segments is preferred.

Infectious virions do not systematically differ in any observable way from replication-defective virions, nor do they differ in a physical way that would allow for their mechanical or chemical separation. For this reason, methods to count infectious virions are based on counting the infections they cause, rather than the particles themselves. In practice, however, not all infection-competent virions contained in a sample will go on to successfully cause infection. Certain experimental conditions, such as temperature or acidity of the medium, can hasten the rate at which virions that were infection-competent in the sample lose infectivity before they can cause infection. This is why, hereafter, we will refer to the quantity measured by infectivity assays as the \emph{infection concentration} or the number of infections the sample will cause per unit volume, rather than its concentration of infectious virions, which is not a measurable quantity. Two main types of assays are used to quantify the infection concentration within a virus sample: (1) the plaque forming and focus forming assays; and (2) assays we will collectively refer to as endpoint dilution (\ID) assays\footnote{Technically, the plaque and focus forming assays are also endpoint dilution assays because they rely on the counting of plaques or foci (the endpoint) as a function of dilutions. However, herein, we will refer to them as plaque or foci forming assays rather than endpoint dilution assays.}, which include the 50\% tissue culture infectious dose (\tcid) or cell culture infectious dose (CCID$_{50}$) or egg infectious dose (EID$_{50}$) assays, etc.

\begin{figure}
\centering
\includegraphics[width=0.5\textwidth]{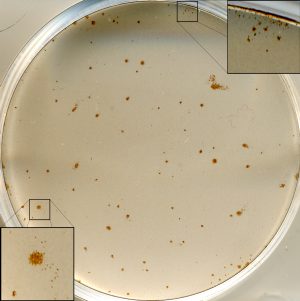}
\caption{\textbf{Examples of challenges in establishing a robust count of infection plaques or foci.} MDCK cells were infected with a sample containing influenza A (H3N2) virions, and cell infection was visualized via staining by antibodies against the matrix (M) viral protein. The uneven liquid distribution along the well's edges means some infectivity is lost or miscounted. It can be hard to distinguish between two merged foci and a single larger uneven focus. It is difficult to determine how small a focus should be counted, and doing so to decide on a focus size threshold to be used consistently for all wells and all samples within a particular experiment. As a result of these difficulties, different individuals will commonly count a different number of foci in the same well. Stained well image graciously provided by Frederick Koster (Lovelace Research Institute, NM, USA).}
\label{fbadplaques}
\end{figure}

The plaque forming assay was introduced by Renato Dulbecco in 1952 \cite{dulbecco1952}, as an improvement over the \ID assay. The plaque forming assay and the focus forming assay, which rely on the same principles, suffer from a number of critical issues that cannot be overcome. For example, the liquid accumulation (meniscus) that forms around well edges means some infectious doses will not get quantified correctly or at all. It can be hard to distinguish two merged plaques from a single large plaque, or to decide how small a plaque one should consider when counting. Some of the difficulties in establishing a robust, unambiguous plaque or focus count for a given well are illustrated in Figure \ref{fbadplaques}. For these reasons, different researchers will count a different number of plaques or foci when observing the same well. This subjectivity in the count means there is opportunity to (sub)consciously count a few more plaques or foci, for example, when expecting a virus strain to be more severe than another or in the absence of an antiviral compound. Ideally, there would be no discretion involved in the counting process of a quantification assay. Indeed, the decision process should be made by a physical, automated measurement, without the possibility of post-facto adjustments of any kind, for any reason.

In contrast, the \ID assay offers a more decisive and robust binary determination as to whether or not infection has taken place in each well (or egg, animal, etc.). This determination is insensitive to small, spatially localized irregularities and is typically unanimously agreed upon by all observers. Therefore, it is less subject to (sub)conscious bias. In fact, this feature of the \ID assay makes it ideal for systematic, machine-based determination of positive wells (or eggs or other culture types), eliminating subjectivity. Furthermore, infection of wells in the \ID assay can be carried out in exactly the same way as planned infection experiments where they will make up the inoculum, e.g., in the same cell type, reproducing whether the inoculum is rinsed or not post-inoculation, and the duration of incubation with the inoculum. In contrast, plaque and focus forming assays can require the use of a semi-solid cellular overlay (e.g., agarose) to restrict the spread of virus beyond cells neighbouring those initially infected by the inoculum. The need to rinse or remove the inoculum to add the semi-solid overlay imposes strict constraints on the timing of this rinse. Because a longer incubation provides more opportunities for infectious virus to cause infection, the number of infectious doses counted via a plaque or focus assay can underestimate the true number of infections that will result when the quantified sample is later used to infect cells under longer incubation periods. The plaque assay can also require the use of different cells than those used in the infection experiments whenever the latter fail to die or detach (form clearly visible plaques) post-infection, making it difficult to predict the number of infections that will result when the quantified sample is later used to infect different cells.

For all its many advantages, the \ID assay currently has one key, remediable weakness: its output quantity, the \tcid\ (or CCID$_{50}$ or EID$_{50}$), does not directly correspond, or trivially relate, to one infectious dose. The simplistic calculations, introduced by \SKlong (\SK) \cite{spearman1908,karber1931} and \RMlong (\RM) \cite{reed1938} nearly a century ago, remain the primary methods to quantify a virus sample's infectivity using the \ID assay. Many research groups rely on spreadsheet calculators that are passed down through generations of trainees or found on the internet, and can contain errors\footnote{For example, versions 2 and 3 of the Excel spreadsheet calculator provided by the Lindenbach Lab at Yale University (\url{http://lindenbachlab.org/resources.html}), which have since been removed.}. While, theoretically, a dose of \unit{1}{\tcid} is expected to cause $-1/\ln(50\%)=\unit{1.44}{infections}$ \cite{bryan1957}, the approximation used by the \SK and \RM methods introduces an often overlooked bias where $\unit{1}{\tcid} \approx \unit{1.781}{infections}$ where $1.781 = \me{\gamma}$ and $\gamma=0.5772$ is the Euler-Mascheroni constant \cite{wulff12,govindarajulu01}. This makes it problematic to experimentally achieve the desired multiplicity of infection when inoculating from a sample quantified via the \SK or \RM methods. Many have proposed replacements for the \RM and \SK calculations with some based on logit or probit transforms of the data \cite{govindarajulu01,labarre01,bryan1957} and others on statistical analysis of the \ID assay output \cite{labarre01,mistry18}. Sadly, none of these improvements were widely adopted, possibly due to a lack of visibility of these publications, or the lack of widespread awareness of the limitations of the \RM and \SK methods.

Thus, the one issue with the \ID assay is not with the assay itself but with the calculation of the \tcid. We submit that for all the reasons outlined above, the \ID assay is experimentally more robust and reliable than the plaque and focus forming assays, and should be preferred over the latter. We propose to:
\begin{enumerate}
\item Continue the use, or encourage the adoption, of the \ID assay (e.g., \tcid\ assay), but to replace its output, the TCID$_{50}$/mL (or CCID$_{50}$/mL, EID$_{50}$/mL, etc.), with a new quantity in units of \textbf{\underline{S}}pecific \textbf{\underline{IN}}fections or \idnew/mL corresponding to the number of infections the sample will cause per mL. The word \emph{specific} highlights the fact that the infectivity of a sample is specific to the particulars of the experimental conditions (temperature, medium, cell type, incubation time, etc.).
\item Replace the \RMlong and \SKlong approximations with a computer software, \newprog (\textbf{\underline{m}}easure of \textbf{\underline{i}}nfectious \textbf{\underline{d}}ose in \textbf{\underline{SIN}}), that relies on Bayesian inference to measure the \idnew/mL of a virus sample. To avoid calculation errors and make the new method widely accessible, \newprog is maintained and distributed as free, open-source software on GitHub (\giturl) for user installation, but also via a free-to-use website application (\progurl) with an intuitive user interface.
\end{enumerate}
Here, we present examples of \newprog being used to analyze influenza and respiratory syncytial virus samples. We demonstrate that \newprog's output, \idnew/mL, is an accurate estimate of the number of infections the sample will cause per unit volume. We show how the accuracy of the \idnew\ concentration estimate is affected by experimental choice of plate layout, including the dilution factor, and the number of replicates per dilution. We compare \newprog's performance to that of the \RM and \SK methods, and demonstrate how the latter estimators are inaccurate under various circumstances, underlining the need to adopt \newprog to quantify virus samples via the \ID assay.

\cleardoublepage
\section{Results}

\subsection{Key features of \newprog's output}
\label{resone}

\begin{figure}
\centering
\includegraphics[width=0.5\textwidth]{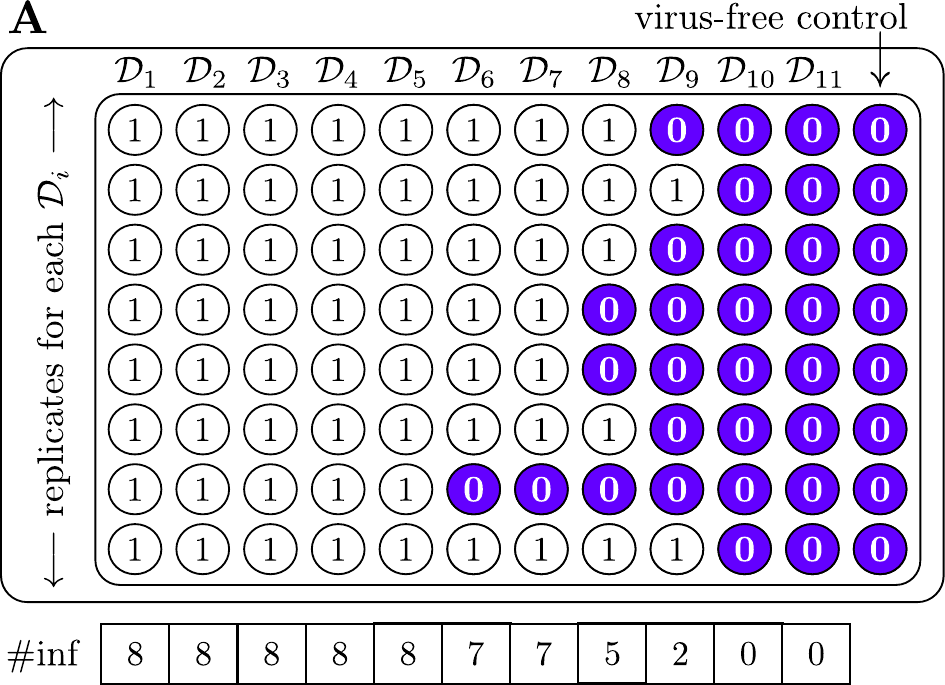}\\[0.5em]
\includegraphics[width=0.67\textwidth]{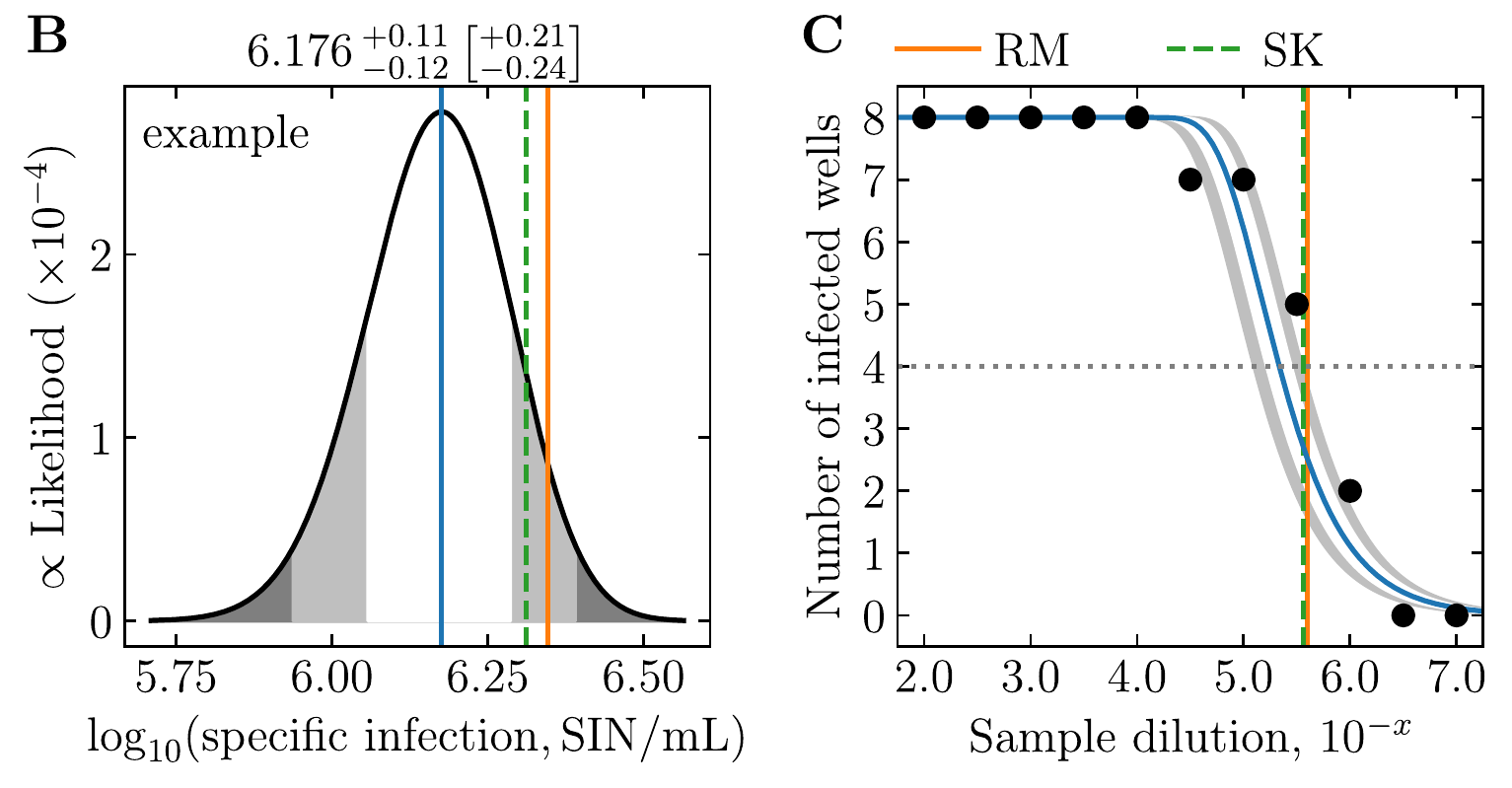}
\caption{%
{\bf Visual representation of \newprog's output for the example \ID plate.}\\
\textbf{(A)} Illustration of the example \ID plate where $\Dil_i$ are the chosen serial dilutions of the sample. For the example described in the text, $\Dil_1=10^{-2}$, $\Dil_2=10^{-2.5}$, ..., $\Dil_{11} = 10^{-7}$, with 8 replicates per dilution. The number of infected wells (\# inf) is indicated at the bottom of each dilution column.
\textbf{(B)} The \newprog-estimated likelihood distribution of the $\log_{10}$ infection concentration, \lidnew, for the example \ID experiment. The vertical lines correspond to \lidnew, based on the most likely value (mode) of \newprog's likelihood distribution (solid blue), or computed from the \RM (solid orange) and \SK (dashed green) approximations of the $\log_{10}(\tcid)$ (see Methods). The $x$-value of the white and light grey region on either sides of the mode indicate the edges of the 68\% and 95\% credible interval (CI), respectively. The \newprog-estimated \lidnew $\text{mode}\pm68\%\left[\pm95\%\right]$ CI are indicated numerically above the graph. %
\textbf{(C)} The number of infected wells (black circles) out of the 8 replicates, as a function of the 11 serial dilutions of the example \ID plate, from the least (leftmost) to the most (rightmost) diluted. For example, $x=3.0$ corresponds to a sample dilution of $10^{-3}$ or 1/1,000. The average (expected) number of infected wells, as a function of sample dilution, is shown for the most likely value of \lidnew\ (blue curve) or its 68\% and 95\% CI (inner and outer edge of the grey bands, respectively). The sample dilution ($x$-value) at which the blue curve crosses the horizontal dotted line (50\% infected wells) corresponds to a concentration of \unit{1}{\tcid} per \ID well volume. The vertical lines indicate the sample dilution that yields a concentration of \unit{1}{\tcid} according to the \RM and \SK approximations.
}
\label{fig:example1}
\end{figure}

Let us consider a fictitious \ID experiment, with 11 dilutions and 8 replicate wells per dilutions, in which the minimum sample dilution, $\Dil_1=1/100=10^{-2}$, is serially diluted by a factor of $10^{-0.5} \approx 0.32$ ($\Dil_2=10^{-2.5}$, $\Dil_3=10^{-3}$, ..., $\Dil_{11}=10^{-7}$), and the total volume of inoculum (diluted virus sample + dilutant) placed in each well is $\Vinoc=\unit{0.1}{\milli\liter}$. Now, consider that a virus sample is measured using this \ID experiment and one observes (8,8,8,8,8,7,7,5,2,0,0) infected wells out of 8 replicates at each of the 11 dilutions, as illustrated in Figure \ref{fig:example1}A.

\newprog provides a graphical output of its results, shown in Figure \ref{fig:example1}B,C for this example. Note how the likelihood distribution for \lidnew (Fig.\ \ref{fig:example1}B) is approximately a normal distribution. This is why $\log_{10}$ of the infection concentration should be used and reported, rather than the concentration itself. \newprog also graphically compares the number of infected wells observed experimentally (Fig.\ \ref{fig:example1}C, black dots) against the theoretically expected values (blue curve and grey CI bands). This graphical representation makes it easy to identify issues with the data entered or with the experiment itself.

Importantly, \newprog provides a more useful quantity to the user than the \tcid: an estimate of the concentration of infections the sample will cause, \idnew/mL. For this example, the concentration is \unit{10^{6.2\pm0.1}}{\idnew/mL}, where 6.2 is the mode (most likely value) of \lidnew, and $\pm0.1$ is its 68\% credible interval (CI). The \idnew/mL corresponds to the number of infections that will be caused per \milli\liter\ of the sample, which can be directly used to determine the sample dilution required to obtain a desired multiplicity of infection (MOI).

\begin{figure}
\centering
\includegraphics[width=\textwidth]{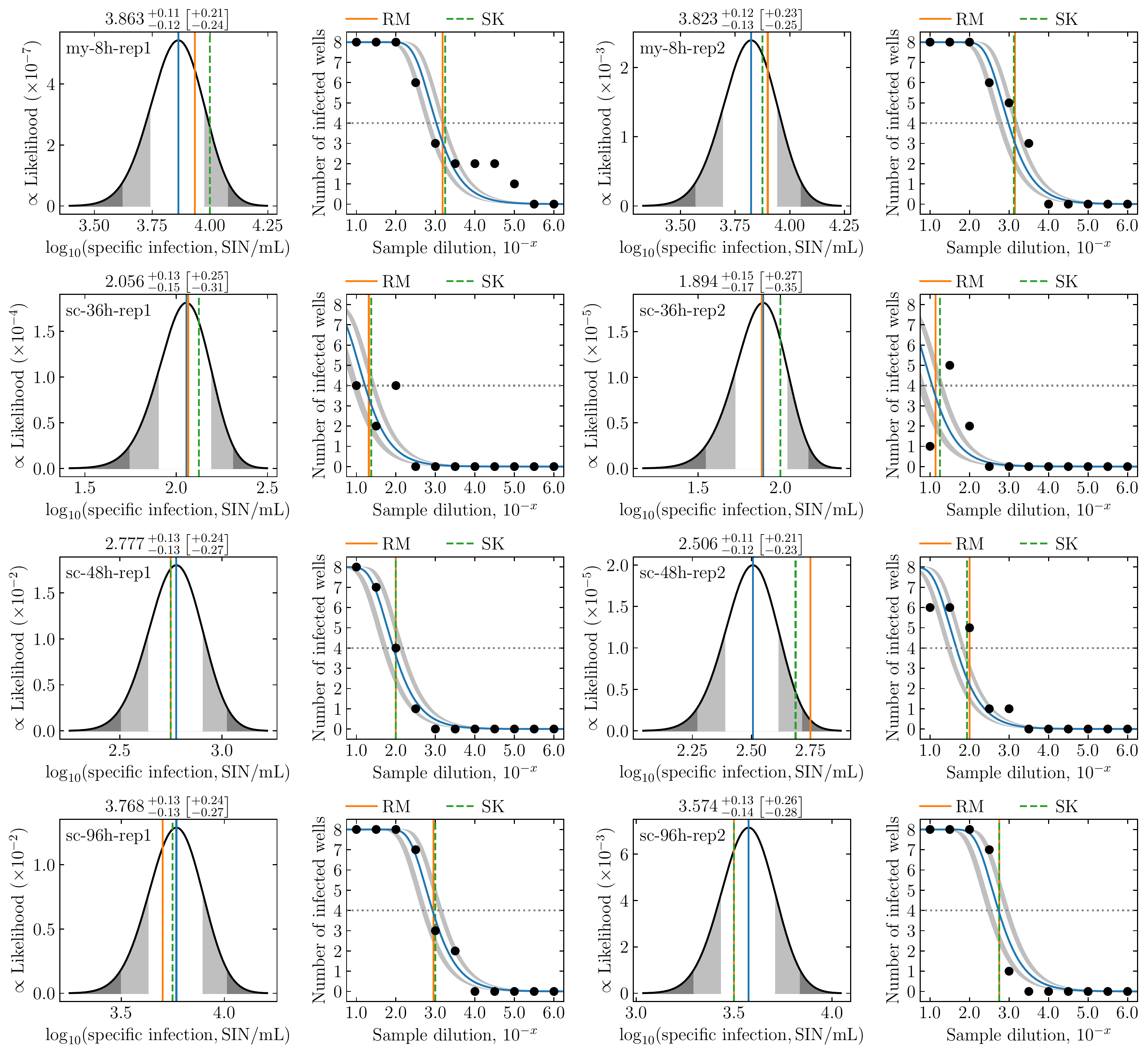}
\caption{%
{\bf Quantification of RSV sampled from \emph{in vitro} infections.}
Each row corresponds to a different experiment (mock-yield [my] or single-cycle [sc]) and sampling time point (e.g., \unit{8}{\hour}, \unit{36}{\hour}), and each sample was measured in duplicate (rep1, rep2). These data were collected from \textit{in vitro} infections with the RSV A Long strain, and were previously reported in \cite{beauchemin19}. The \ID measurement experiment were conducted using a plate layout of 11 dilutions, with 8 replicates per dilution, an inoculum volume of $\Vinoc=\unit{0.1}{\milli\liter}$, serial dilutions from $\Dil_1=10^{-1}$ to $\Dil_{11}=10^{-6}$, separated by a dilution factor of $10^{-0.5}$.
}
\label{fig:rsv-plots}
\end{figure}

In a laboratory setting, \ID experiments can be performed in batches, such as to quantify the infectious concentration in samples collected at several time points over the course of a cell culture infection. For such applications, \newprog provides a comma separated value (csv) template file readily editable in a spreadsheet program, to collect and submit the results for batch processing. Details on the format of the template file are available on \newprog's website (\progurl). Figure \ref{fig:rsv-plots} illustrates the output for a subset of measurements for \emph{in vitro} infection with the respiratory syncytial virus (RSV). Each sample was measured twice, and \newprog's estimates are in good agreement with one another (within 95\% CI).

The $y$-axis in the left graph panels of \newprog's graphical output is the non-normalized scale of the likelihood distribution for \lidnew, which ranges between $10^{-7}$ and $10^{-2}$. The scale loosely relates to the likelihood of observing a particular \ID experimental outcome (see Methods). Unlikely \ID outcomes appear as large departures of the observed number of infected wells (right panels, black dots) from what is theoretically expected (right panels, curve). It is interesting that the uncertainty (CI) of \newprog's estimated \lidnew appears to be independent of how much the \ID outcome deviates from theoretical expectations. That is, the accuracy of \newprog\ is not strongly affected even when it is provided more unlikely, noisy experimental data. This robustness is explored further below.

\subsection{Comparing \idnew\ to \tcid\ and \pfu\ virus sample concentrations}

\begin{figure}
\includegraphics[width=\linewidth]{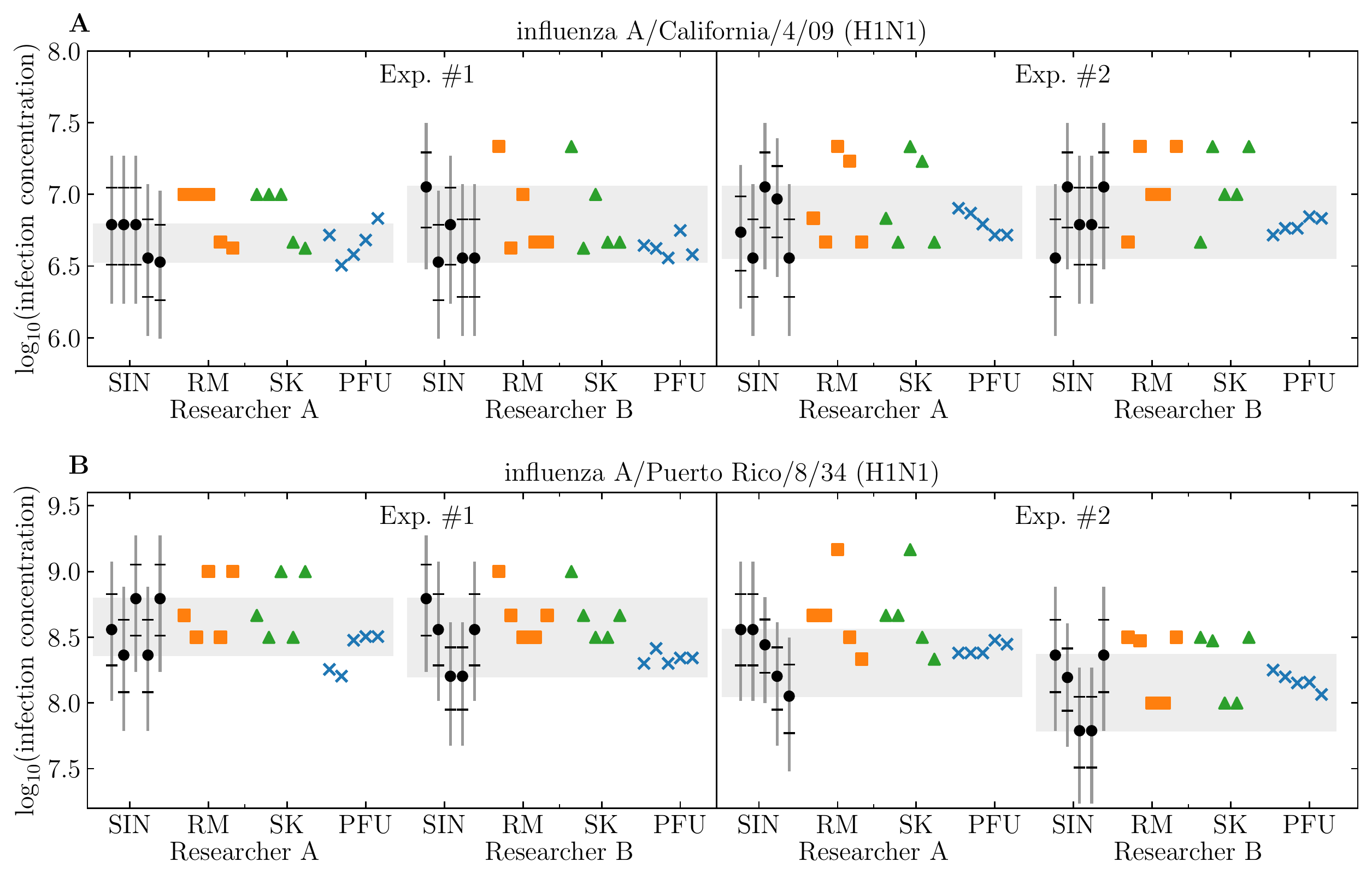}
\includegraphics[width=\linewidth]{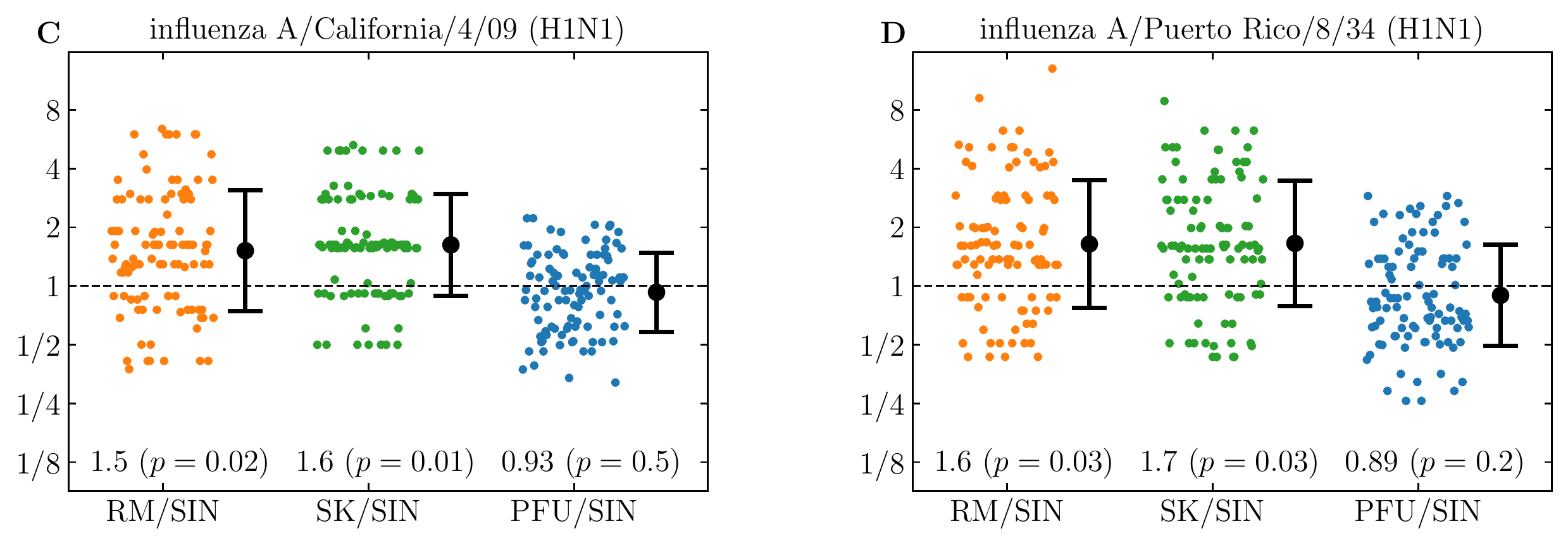}
\caption{%
{\bf Comparing \idnew\ to \tcid\ and \pfu\ for influenza A virus samples.} %
\textbf{(A,B)} The infection concentration in two influenza A (H1N1) virus strain samples was measured via both an \ID assay and a plaque assay (x, PFU). The \ID assay was quantified in $\log_{10}(\tcid)$ using the \RM (square) or \SK (triangle) methods, or in $\log_{10}(\idnew)$ using \newprog (circle with 68\%,95\% CI). Each of the 2 strain samples was measured over 2 separate experiments (Exp.\ \#1, \#2), performed each time by 2 different researchers (Researcher A or B), with 5 biological replicates each. The grey bars indicate the range of $\log_{10}(\idnew)$ values across the 5 replicates. The RM, SK, and SIN measures were estimated for each replicate based on the same \ID plate. The experimental details are provided in Methods. %
\textbf{(C,D)} The $\log_{10}$ of the ratio between either the \tcid\ via the RM or SK method or the \pfu, over the \idnew\ via \newprog. The ratios were computed for each replicate ($5\times5$ replicates), per experiment, per researcher (25 replicates $\times$ 2 researchers $\times$ 2 experiments = 100 ratios) shown as individual symbols (dots) for each method (RM, SK, \pfu). The mean and 68\% CI of the 100 ratios are indicated numerically and as black circles with error bars.
}
\label{fig:amber-pfu}
\end{figure}

The \newprog calculator provides an estimate of the number of infections that will be caused per mL of a virus sample (\idnew/mL). In principle, a plaque assay also measures the number of infections a sample will cause, with each infection expected to develop into a plaque. If a plaque assay is performed under experimental conditions and protocols as similar as possible to those of the \ID assay (i.e., using the same cells, medium, period of incubation, rinsing method, etc.), \newprog's \idnew/mL estimate is expected to be comparable, in theory, to the number of \pfu/mL observed in the plaque assay. In practice, however, the plaque assay likely provides a biased estimate of the concentration of infections in a sample due to its many experimental issues, discussed in the Introduction. To evaluate \newprog's performance compared to existing methods, the infection concentration in two influenza A (H1N1) virus strain samples were measured via both plaque and \ID assays, and their concentration in units of \pfu, \tcid, and \idnew\ were compared (Fig.\ \ref{fig:amber-pfu}). Details regarding the samples, and how the plaque and \ID assays were performed are provided in Methods.

The \tcid\ concentrations estimated via the \RM and \SK methods are $\sim$1.5--1.7 times larger (Fig.\ \ref{fig:amber-pfu}C,D) than the \idnew\ concentration, and the set of ratios are statistically inconsistent with the assumption of equality ($p$-value: 0.01--0.03). Theoretically, \unit{1}{\tcid} is expected to cause \unit{1.44}{infections} ($=1/\ln(2)$) \cite{bryan1957}. However, the \RM or \SK approximations are known to introduce a bias such that \unit{1}{\tcid} estimated by these methods is expected to cause \unit{1.781}{infections} ($=\me{\gamma}$ where $\gamma=0.5772$ is the Euler-Mascheroni constant) \cite{wulff12,govindarajulu01}. Using the \RM, \SK, and \idnew\ measurements presented in Figure \ref{fig:amber-pfu}A,B, we confirmed%
\footnote{The mean $\log_{10}(\mathrm{ratio})$ was re-computed for $\mathrm{ratio}=(\mathrm{\RM}/1.781)/\idnew$ and (\SK/1.781)/\idnew, and found to be 0.85--0.93. This is statistically consistent ($p$-value: 0.1--0.3) with the assumption of equality, i.e., $\mathrm{ratio} = 10^0 = 1$.} %
 that $\unit{1.781}{\idnew} \approx \unit{1}{\tcid}$ when the latter is estimated via the \RM or \SK approximations, as expected theoretically if \idnew\ is indeed measuring the infection concentration in a sample.

Similarly, the ratio of the \pfu\ concentration determined via the plaque assay and the \idnew\ concentrations estimated by \newprog is $\sim$0.89--0.93, which is statistically consistent with the assumption of equality ($p$-value: 0.2--0.5). These results confirm the theoretical expectation that $\unit{1}{\pfu} \approx \unit{1}{\idnew}$ when the plaque and \ID assays are performed in the same manner, as was the case here. This provides further support, via two independent assays, that the \idnew\ concentration estimated by \newprog from the \ID assay is a robust measure of the infection concentration of a virus sample.

\subsection{Comparing \newprog's performance to that of the \RM and \SK methods}

\begin{figure}
\includegraphics[width=\linewidth]{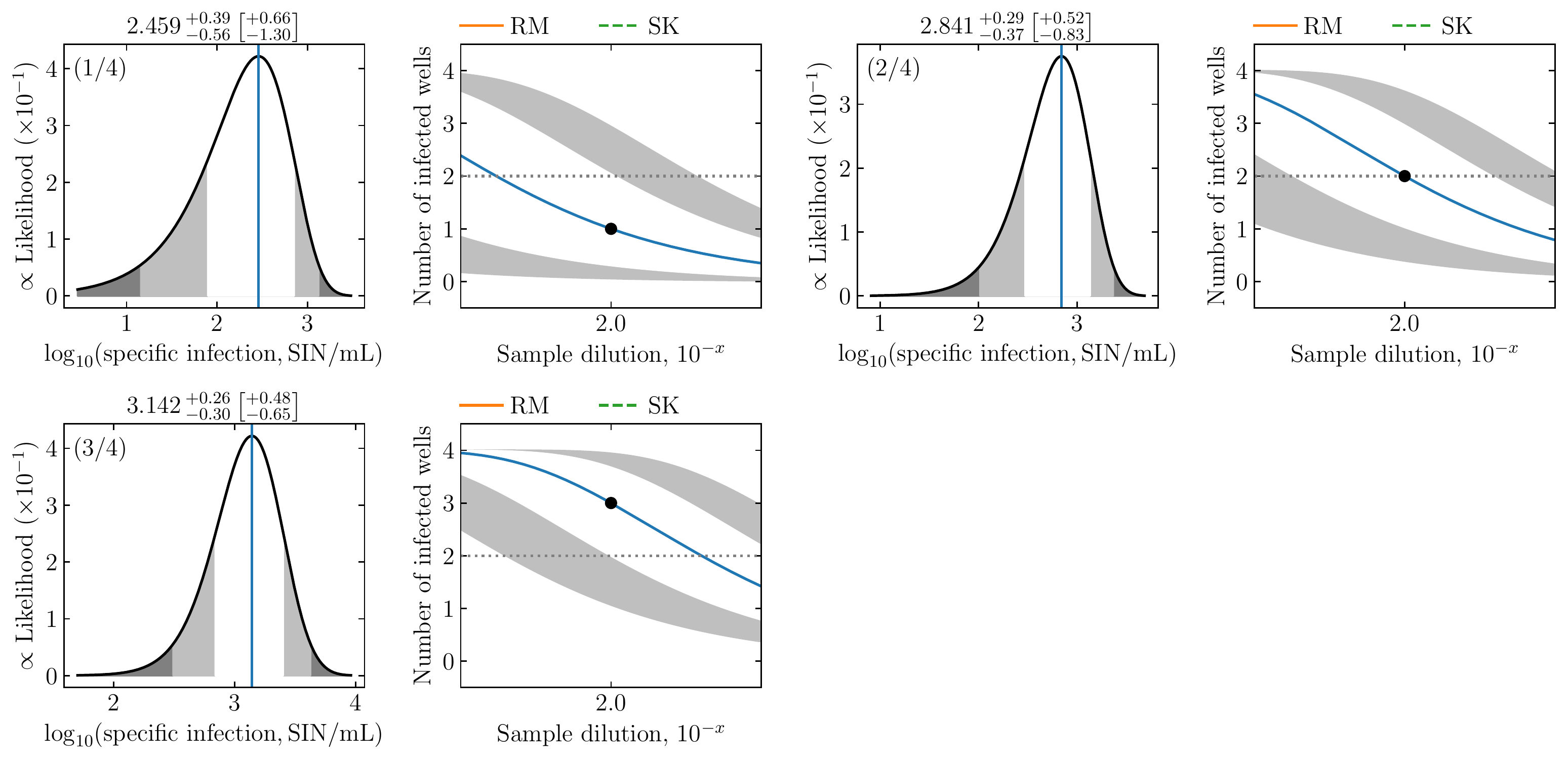}
\caption{%
{\bf \newprog's estimate of a sample's infection concentration based on a single dilution.} %
This is a simulated example of an \ID plate with an inoculation volume of $\Vinoc=\unit{0.1}{\milli\liter}$. Instead of serial dilutions, a single dilution ($\Dil_1=0.01$) is used, and either 1, 2 or 3 well(s) out of the 4 replicate wells are infected. As the fraction of infected wells increases, the uncertainty on the estimate (68\% and 95\% CIs) decreases, and the likelihood distribution becomes more symmetric (Normal-like).
}
\label{fig:onecol}
\end{figure}

The \RM and \SK methods rely on the number of infected wells decreasing as dilution increases. Their estimates are affected when the number of infected wells remains unchanged or even increases as dilution increases, which statistics tell us can reasonably occur experimentally. The \RM and \SK methods also mostly require that at the lowest and highest sample dilutions, all wells be infected and uninfected, respectively. In contrast, \newprog is robust to these issues. Figure \ref{fig:onecol} demonstrates how \newprog can provide an estimate for the \lidnew\ in a sample using the number of infected wells at a single dilution, as long as at least one well is uninfected if all others are infected or vice-versa. This is because \newprog relies on Bayesian inference, i.e., when more than one column is available, it uses information from each column successively to revise and improve its estimate. This allows \newprog to correct for even large deviations from theoretical expectations, and thus improves its accuracy.

\begin{figure}
\centering
\includegraphics[width=\textwidth]{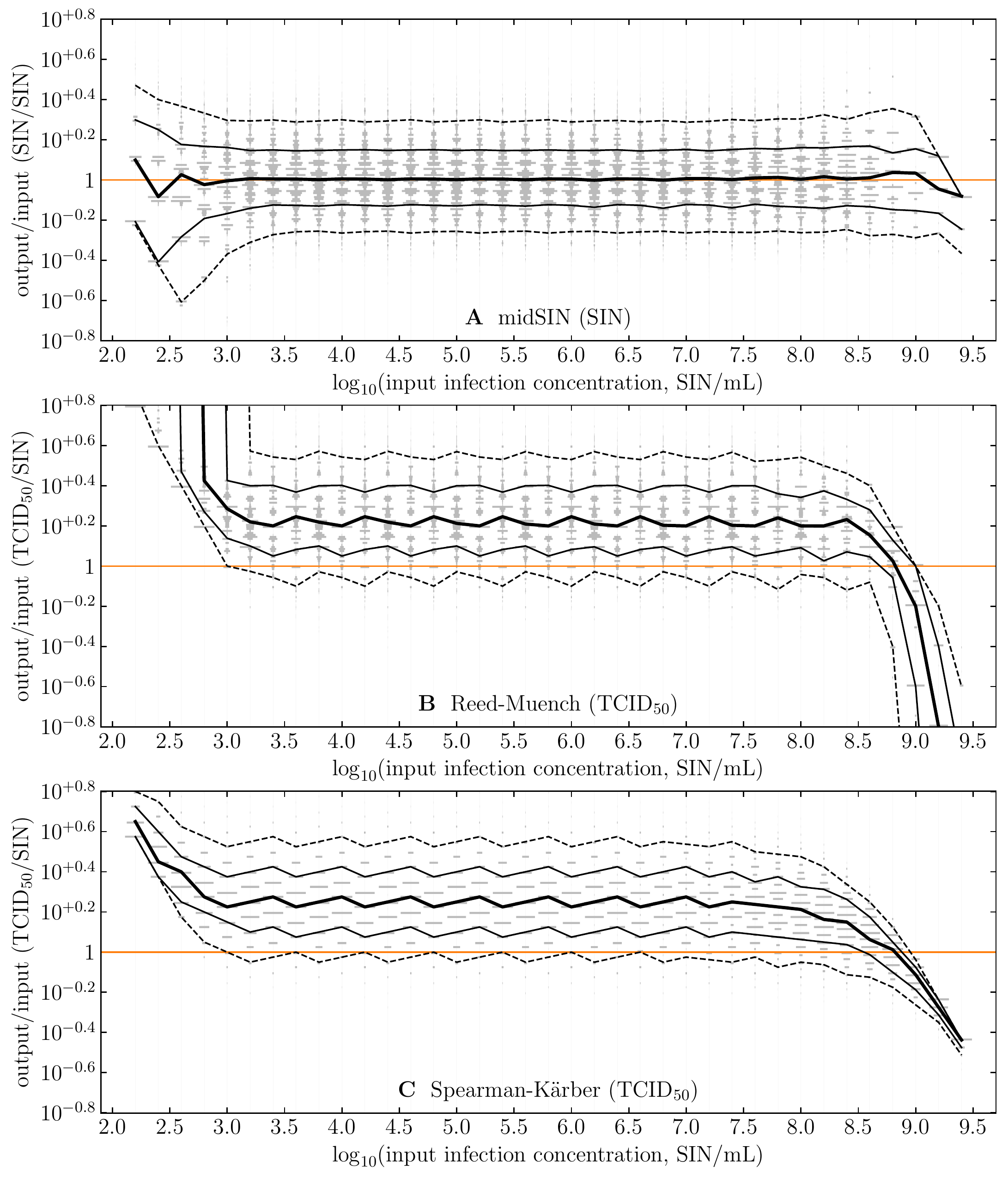}
\caption{%
{\bf Comparing known input to estimated output concentrations.} %
For each input concentration between $10^{2.2}$ and $10^{9.4}$, one million random \ID experiment outcomes (\# of positive wells in each dilution column) were generated. For each \ID outcome, either (A) \newprog was used to determine the most likely \lidnew; or the (B) \RM or (C) \SK method was used to estimate the $\log_{10}(\tcid/\milli\liter)$. Vertically stacked grey bands at each input concentration are sideways histograms, proportional to the number of \ID outcomes that yield a given $y$-axis value. The black curves join the median (thick), 68$^\text{th}$ (thin) and 95$^\text{th}$ (dashed) percentile of the histograms, determined at (but not between) each input concentration. A plate layout of 11 dilutions, with 8 replicates per dilution, an inoculum volume of $\Vinoc=\unit{0.1}{\milli\liter}$, serial dilutions from $\Dil_1=10^{-2}$ to $\Dil_{11}=10^{-8}$, separated by a dilution factor of $10^{-0.6}\approx 1/4$ were used in the simulated \ID experiments.
}
\label{fig:inouterr}
\end{figure}

Figure \ref{fig:inouterr} illustrates how well the \newprog, \RM, and \SK methods recover a known input sample concentration in simulated \ID experiments, based on a plate layout consisting of 11 dilutions ($\Dil_1=10^{-2}$ to $\Dil_{11}=10^{-8}$), a dilution factor of $1/4$, and 8 replicates per dilutions. The infection concentration estimated by \newprog is in excellent agreement with the input concentration. For the \RM and \SK methods, which estimate the $\log_{10}(\tcid/\milli\liter)$ rather than the \lidnew, the agreement is generally poor due to the bias they introduce. Furthermore, the \RM and \SK predictions are more variable (wavy pattern), and lose accuracy dramatically as the sample concentration approaches the limits of detection (the 2 ends) which, for the example plate layout simulated here, is around \unit{10^3}{\idnew/\milli\liter} and \unit{10^9}{\idnew/\milli\liter}. Interestingly, the basic calculations behind the \RM and \SK methods constrain the set of values they can return (sparsely populated grey histograms), compared to the more continuous range returned by \newprog, which contributes to its increased accuracy.

\subsection{Estimate accuracy as a function of plate layout}

\begin{figure}
\centering
\includegraphics[width=\textwidth]{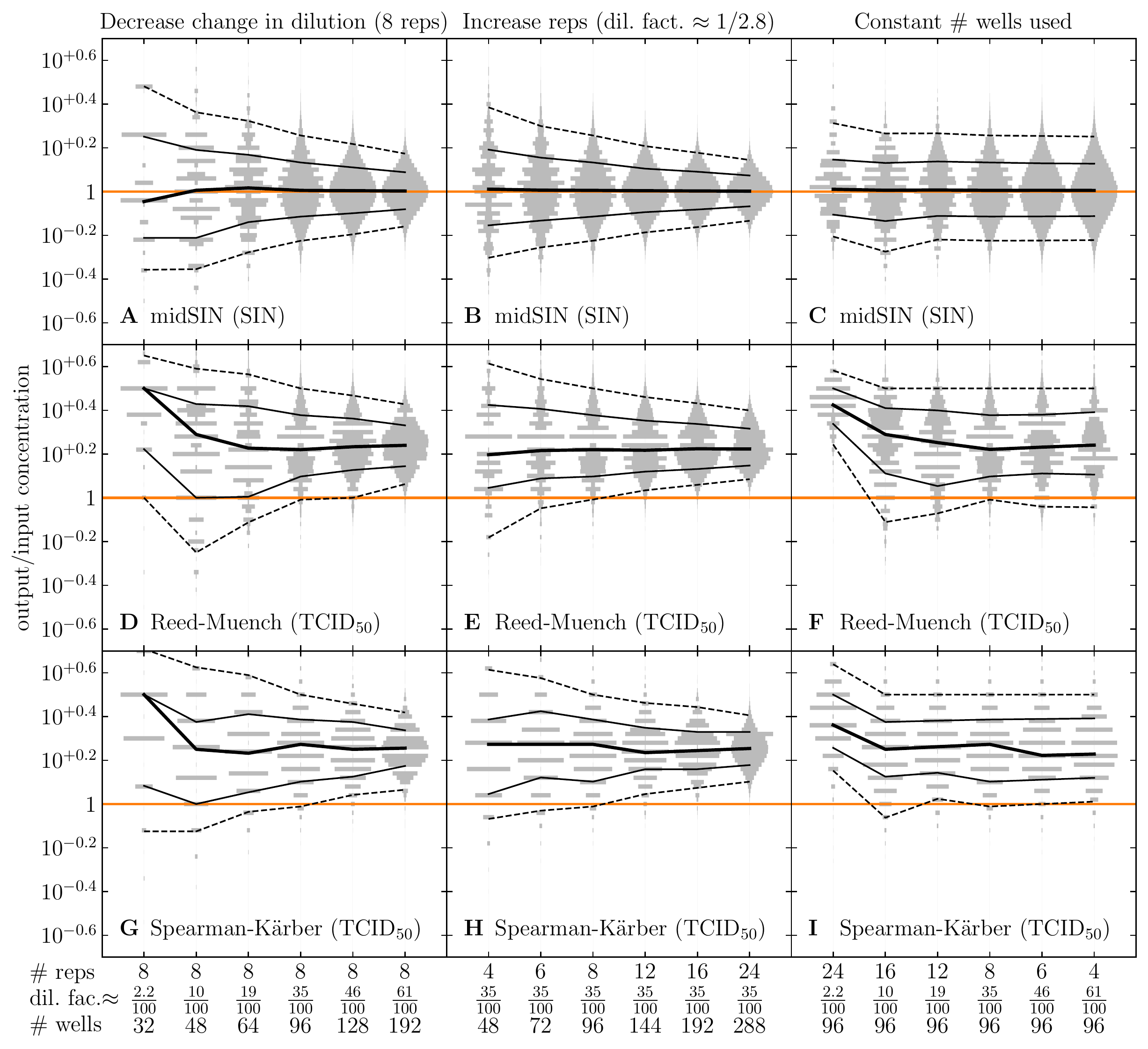}
\caption{%
{\bf Comparing the effect of the dilution factor and number of replicates per dilution.} %
The effect of either (A,D,G) decreasing the change in dilution (from a dilution factor of $2.2/100$ to $61/100$) while keeping 8 replicates per dilution; or (B,E,H) increasing the number of replicates per dilutions (4 to 24) while keeping a fixed dilution factor ($\approx35/100$); or (C,F,I) increasing the dilution factor while decreasing the number of replicates, keeping a fixed number of 96 wells used in total to titer one virus sample. Different rows represent the ratio of the estimated output concentration using (A--C) \newprog in \idnew/mL, (D--F) \RM or (G--I) \SK in \tcid/mL, and the input concentration. In all cases (A--I), the input concentration was \unit{10^5}{\idnew/mL}, and as the dilution factor was varied, the highest and lowest dilutions in the simulated \ID plate were held fixed to $\Dil_1=10^{-2}$ and $\Dil_\text{last}=10^{-7}$, respectively, by changing the total \# of dilutions performed (simulated). Everything else is generated or computed as described in the caption of Figure \ref{fig:inouterr}.
}
\label{fig:plate-layout}
\end{figure}

In Figure \ref{fig:rsv-plots}, we observed that even for large discrepancies between the expected (right panels, blue curve) and observed (right panels, black dots) \ID assay outcome, the uncertainty (CI) of \newprog's estimate remains relatively unchanged. This apparent robustness is because the uncertainty is primarily determined by the experimental design, namely the change in dilution between columns (dilution factor) and the number of replicate wells per dilution. Figure \ref{fig:plate-layout} explores the impact of varying either only the dilution factor, or only the number of replicates at each dilution, or varying one at the expense of the other by using a fixed number of wells (96 wells). When using \newprog, smaller changes in dilution (e.g., going from a dilution factor of 2.2/100 to 61/100) or more replicates per dilution (4 to 24) each improves the measure's accuracy (narrower CIs) by comparable amounts, but only when the total number of wells is allowed to increase to accommodate the change. When the total number of wells used is fixed, changing one at the expense of the other leaves the accuracy (CI) unchanged. This is somewhat also true for the $\log_{10}(\tcid)$ output concentration estimated by the \RM and \SK methods. However, at the smallest dilution factors (10/100 and 2.2/100), the bias introduced by the \RM and \SK methods becomes even larger and more unpredictable. For the input concentration considered in Figure \ref{fig:plate-layout} (\unit{10^5}{\idnew/mL}), the dilution at which 50\% of wells are infected is near the middle dilution. For sample concentrations such that 50\% infected wells occur near or at the lowest or highest dilution chosen, the effect is even more significant.

Figure \ref{fig:plate-layout} also demonstrates that varying the dilution by smaller increments (e.g., a dilution factor of 61/100 rather than 10/100) provides greater granularity (uniqueness) of \ID plate outcomes, and thus, greater accuracy of the \lten infection concentration estimates. Here, a distinct plate outcome means a distinct number of infected wells at each dilution, with no distinction as to exactly which of the replicate wells (e.g., the second versus the fourth) is infected at each dilution. An \ID plate with serial dilutions ranging over 6 orders of magnitude (e.g., $10^{-2}$ to $10^{-7}$), with 4 different dilutions and 24 replicates/dilution (i.e., dilution factor of 2.2/100) provides $\sim10^6$ ($[24+1]^{4}$) possible, distinct \ID plate outcomes. In contrast, a plate with the same serial dilution range, but with 24 different dilutions and 4 replicates/dilution (i.e., dilution factor of 61/100) yields $\sim10^{17}$ ($[4+1]^{24}$) distinct outcomes. More generally, $[\text{reps}+1]^\text{dils}$ is the number of distinct plate outcomes for a chosen number of dilutions (dils) and replicates (reps). Having fewer possible plate outcomes means that a larger range of concentrations would share the same most-likely \ID plate outcome, yet each plate outcome only maps to one (the most likely) concentration estimate. This means that with fewer dilutions, the concentration estimate is forced to take on the nearest possible value it can take (Fig.\ \ref{fig:plate-layout}, the next grey bar), and the accuracy of the concentration estimate is therefore reduced. So although having a greater number of dilutions is more labour intensive, it should be preferred over having a greater number of replicates per dilution.

\cleardoublepage
\section{Discussion}

We have introduced a new calculator tool called \newprog to replace the \RMlong (\RM) and \SKlong (\SK) calculations to quantify the infectivity of a virus sample based on a \tcid\ endpoint dilution (hereafter \ID) assay. Rather than estimating the \tcid\ of a virus sample, \newprog calculates the number of infections the sample will cause, reported in units of specific infections (\idnew). It does so without requiring any changes to current \ID assay protocols, and can be accessed for free via an open-source web-application (\progurl). Importantly, since the \idnew\ of a virus sample corresponds to the number of infections it will cause, it can be used directly to determine what dilution of the sample will achieve the desired multiplicity of infection (MOI).

We showed that \newprog provides more accurate and robust estimates than the biased \RM and \SK approximations. We confirmed that the \RM and \SK approximations overestimate the \tcid\ by 23.5\%, such that \unit{1}{\tcid} estimated by these methods will cause 1.781 rather than \unit{1.44}{infections} \cite{wulff12,govindarajulu01}. While in theory one can obtain the intended MOI by multiplying the \tcid\ by 0.7 (or rather $\ln(2)=0.693$), one should instead multiply by 0.561 to account for the overestimation by \RM and \SK. Even when accounting for the overestimation, we showed that these methods perform particularly poorly when too few replicate wells per dilutions are used or when the change in dilution is large between successive serial dilutions. The two methods perform especially poorly when quantifying samples whose infection concentration approaches, but is still well within, the detection limit of the \ID assay. In such cases, the bias introduced by these methods becomes even larger and more significant. For example, if the minimum and maximum dilutions of an \ID plate are $10^{-2}$ and $10^{-8}$, virus samples with a concentration less than \unit{10^{2.2}}{\idnew} or greater than \unit{10^{7.6}}{\idnew} per inoculated well volume (typically \unit{0.1}{mL}), will see their concentration estimated with an even larger bias by the \RM and \SK methods.

Using \newprog, rather than \RM or \SK, to measure the infectivity of a virus sample based on an \ID assay does not require any change to \ID experimental protocols and methods currently in use in one's laboratory (e.g., dilution factor, replicate per dilution, minimum dilution). Indeed, we demonstrated that \newprog can estimate a virus sample's \idnew\ concentration based on even just a single dilution, as long as only a fraction of the replicate wells are infected at that dilution. For a given number of \ID wells used to titrate the sample and fixed minimum and maximum dilutions (\ID detection range), we showed that having smaller changes between dilutions (a larger number of serial dilutions) is better than having more replicates per dilution. So those wishing to improve the accuracy in estimating the infectivity of their virus samples should consider using more wells in titrating each virus sample, and favouring smaller dilution changes over more replicates. For example, using 11 dilutions, with a 4-fold dilution factor between dilutions and 8 replicate wells per dilution uses up 88 wells, leaving 8 wells of a 96-well plate for controls. This \ID plate design, analyzed using \newprog, accurately measures virus sample concentrations ranging over $\sim$6 orders of magnitude (e.g., [$10^{1}$--$10^{7}$]\,\idnew/mL, or [$10^{6}$--$10^{12}$]\,\idnew/mL, etc.) with an accuracy of $\sim$1.6-fold ($\times 10^{\pm 0.2}$, 95\% CI). In comparison, using 7 dilutions, with a 10-fold dilution factor, and 4 replicates (which uses 28 rather than 88 wells) would also span 6 orders of magnitude, but with an accuracy of $\sim$3.2-fold ($\times 10^{\pm 0.5}$, 95\% CI). To put these 2 accuracies in perspective: \unit{1}{\milli\liter} of a sample measured to contain \unit{10}{\idnew/mL}, is expected to yield either 6--16 or 3--31 infections 95\% of the time, given an accuracy of either $\times 10^{\pm 0.2}$ or \unit{\times 10^{\pm 0.5}}{\idnew/mL}, respectively. Such an important decrease in accuracy means a reduced ability to detect experimental changes as statistically signficant, with the $\times10^{\pm0.5}$ accuracy requiring a $>$10-fold change for statistical significance. Failing to identify a change as statistically significant as part of a study is far more costly than using a few more wells for each sample to increase measurement accuracy, and thus the statistical power of the study.

The \newprog-estimated \idnew\ obtained from an \ID assay was also compared to the \pfu\ from a plaque assay for a set of influenza A virus samples. When the plaque and \ID assays are performed as identically as possible (cell type, incubation time, etc.), as was the case here, $\unit{1}{\idnew}\approx\unit{1}{\pfu}$. This demonstrates that indeed \newprog's \idnew\ is a measure of the number of infections a virus sample will cause. However, as mentioned, the plaque and focus forming assays often impose experimental requirements (e.g., an early rinse of the inoculum to add agarose, use of cells with pronounced CPE). Such constraints on the plaque or focus assay inoculation protocol make it nearly impossible to relate the number of plaques or foci observed to the number of infections the virus sample will cause under the intended, experimental infection conditions (e.g., late or no inoculum rinse, no agarose, to infect cells exhibiting no significant CPE). Adding to this the subjectivity of counting plaques or foci, it is clear the \ID assay combined with \newprog to estimate the \idnew\ concentration of a virus sample is more accessible, accurate, and predictive.

Beyond the work presented herein, the development of \newprog will continue online, as we implement new features and inputs for integration with various colorimetric and fluorescence instruments. The ease of use of \newprog and the greater usefulness and relevance of \idnew\ as a measure of a virus sample's infectivity make them far superior to all currently available alternatives, including the PFU, FFU, \tcid, and other \idfif measures. We hope to see them adopted widely.

\cleardoublepage

\subsection*{Acknowledgements}

The authors wish to thank Frederick Koster (Lovelace Respiratory Research Institute, NM, USA) for providing the antibody stained well image, and Evan Williams (UTHSC, UT, USA) for technical assistance.

\subsection*{Funding}

This work was supported in part by Discovery Grant 355837-2013 (CAAB) from the Natural Sciences and Engineering Research Council of Canada (\url{www.nserc-crsng.gc.ca}), Early Researcher Award ER13-09-040 (CAAB) from the Ministry of Research and Innovation of the Government of Ontario (\url{www.ontario.ca/page/early-researcher-awards}), by the Interdisciplinary Theoretical and Mathematical Sciences programme (iTHEMS, \url{ithems.riken.jp}) at RIKEN (CAAB), and by R01 AI139088 (AMS, APS, LCL) from the NIH NIAID (\url{www.niaid.nih.gov}). The funders had no role in study design, data collection and analysis, or decision to publish.

\subsection*{Authors contribution}

CAAB was responsible for study conceptualization, and project administration, DC, DCW, CQ, CAAB all contributed to the development of the methodology, data analysis, software, visualization. AMS, APS, LCL carried out experimentation, CAAB, AMS were responsible for funding acquisition, experimental planning, and supervision. DC and CAAB contributed to the original manuscript draft, and all authors contributed to its review and editing.

\cleardoublepage
\section{Methods}

\subsection{The mathematics of the dose-response assay}

\subsubsection{Considering a single well}

Consider a virus sample of volume $\Vtot$ which contains an unknown concentration of infectious virions, $\Cvir$, which we aim to determine. Drawing a small volume, $\Vinoc<\Vtot$, from the sample of volume $\Vtot$, is analogous to drawing balls out of a bag containing green and yellow balls, and considering green balls a success, and yellow ones a failure. It is a series of Bernoulli trials where
\begin{description}
\item[$n=\Vinoc/\Vvir$] is the number of draws, i.e., the number of virion-size volumes (\Vvir) drawn from the sample to form the inoculum volume (\Vinoc), analogous to the number of balls drawn.
\item[$k$] is the number of successes, i.e., the number of infectious virions drawn from the sample to form the inoculum, analogous to the number of green balls drawn.
\item[$p$] is the probability of success, i.e., the fraction of virion-size volumes in the sample that are occupied by infectious virions, analogous to the probability of drawing a green ball.
\end{description}
The probability of success, $p$, is related to the concentration of infectious virus in the sample, \Cvir, as
\be
p = \frac{\text{Number of virions in sample}}{\text{Number of virion-size volumes in the sample}} = \frac{\Cvir\Vtot}{\Vtot/\Vvir} = \Cvir\Vvir\ ,
\ee
where \Cvir is the quantity we aim to estimate. Unlike the ball analogy where it is easy to count how many green balls $k$ were drawn, after having drawn $n$ virion-size volumes from the sample into our inoculum, we cannot count how many infectious virions were drawn into the inoculum. However, if this inoculum is deposited onto a susceptible cell culture, we can observe whether or not infection occurs, and this would indicate that the inoculum contained at least one or more infectious virions. Note that, as explained in the Introduction, even a productively infectious virion, i.e., one capable of completing the full virus replication from attachment to progeny release, might not result in a productive infection. As such, from hereon, $\Cvir$ is used to designate the concentration of specific infections in the sample, which is smaller or equal to the concentration of infectious virions, i.e., measures a subset of the infectious virions.

Having deposited the inoculum into one well of the 96-well plate of our \ID experiment, the likelihood that the well will \emph{not} become infected corresponds to the likelihood of having drawn $k=0$ infectious virions (or rather, specific infections) out of the $n$ virion volumes that make up our inoculum, namely
\begin{align}
\qnoinf &= \text{Binomial}(k=0|n=\Vinoc/\Vvir,p=\Cvir\Vvir) \label{eqn:binomqnoinf} \\
&= \frac{n!}{0! (n-0)!} p^0 (1-p)^{n-0} = (1-p)^n \nonumber \\
\qnoinf &= \left(1-\Cvir\,\Vvir\right)^{\Vinoc/\Vvir} \nonumber
\end{align}
where \qnoinf can be simplified by realizing that
\begin{align*}
\ln(1-x) &\stackrel{|x|<1}{=} -x - \frac{x^2}{2} - \frac{x^3}{3} - ... \stackrel{|x|\ll1}{\approx} -x \\
\ln(\qnoinf) &= \frac{\Vinoc}{\Vvir}\ \ln(1-\Cvir\,\Vvir) \approx \frac{\Vinoc}{\Vvir}\ (-\Cvir\,\Vvir) = -\Cvir\,\Vinoc \ .
\end{align*}
As such,
\begin{equation}
\qnoinf = (1-\Cvir\,\Vvir)^{\Vinoc/\Vvir} \approx \exp\left[-\Cvir\,\Vinoc \right]
\label{eqn:qnoinf}
\end{equation}
where \qnoinf and $(\Cvir\Vvir)\in[0,1]$ because $\Cvir=\Nvir/\Vtot$ and the number of specific infections in the sample, \Nvir, is at a minimum zero, and at most the maximum number of virion-size volumes that can physically fit in the sample volume, namely $\Vtot/\Vvir$. As such, the maximum possible infection concentration, given a sample of volume \Vtot, is $\Cvir=(\Vtot/\Vvir)/\Vtot = 1/\Vvir$, and $\Cvir\in[0,1/\Vvir]$.

\subsubsection{Considering replicate wells at a given dilution}

The \ID assay is based on serial dilutions of the sample, with each dilution separated by a fixed dilution factor. We define the dilution factor $\in(0,1)$ as the fraction of the inoculum volume drawn from the previous dilution. For example, if the inoculum for a well, $\Vinoc=\unit{100}{\micro\liter}$, comprises \unit{10}{\micro\liter} drawn from the previous dilution and \unit{90}{\micro\liter} of dilution media, the dilution factor is $10/100=0.1$. If the serial dilution begins with a dilution of $\Dil_1 = 0.2$, then the following dilution will be $\Dil_2 = 0.02$. In Eqn.\ \eqref{eqn:binomqnoinf}, the dilution under consideration, $\Dil_i$, will affect $n$, the number of virion-sized volumes drawn from the sample and deposited into the wells of the $i^\mathrm{th}$ dilution, such that $n=\Dil_i\Vinoc/\Vvir$. Therefore, the probability that a well at the $i^\mathrm{th}$ dilution will \emph{not} become infected is given by
\begin{equation}
\qi \equiv \qnoinf^{\Dil_i} = (1-\Cvir \Vvir)^{\Dil_i\Vinoc/\Vvir} \approx \exp \left[ -\Cvir \Vinoc \Dil_i \right]
\label{eqn:Pi}
\end{equation}
where $1-\qi$ is the probability of infection for a well at the $i^\mathrm{th}$ dilution, where $\Dil_i\in[0,1]$.

When conducting an \ID assay, each dilution in the assay contains a number of independent infection wells (replicates), all inoculated with the same dilution, $\Dil_i$. This is analogous again to drawing balls out of a bag, but this time there are $n_i$ draws (replicate wells), and the probability of success (i.e., that a well becomes infected) is simply one minus the probability of failure (i.e., that a well does not become infected, \qi). The probability that $k_i$ out of the $n_i$ wells become infected at dilution $\Dil_i$, is described by the Binomial distribution
\be
\text{Binomial}(k=k_i|n=n_i, p=1-\qi) = \frac{n_i!}{k_i! (n_i-k_i)!}\, (1-\qi)^{k_i}\, \qi^{n_i-k_i} \propto (1-\qnoinf^{\Dil_i})^{k_i}\, \qnoinf^{\Dil_i(n_i-k_i)}
\ee
where $n_i$ is the number of replicate wells at each dilution, but could be less if any well at dilution $\Dil_i$ are spoiled or contaminated.

However, our interest is not in determining $k_1$ given \qnoinf, but rather in determining \qnoinf given that we observed $k_1$ infected wells out of $n_1$ wells in the first column. To this aim, we can make use of Bayes' theorem which, in our context, can be expressed as
\be
\Pcal(p|\text{data}) = \frac{\Pcal(\text{data}|p)\ \Pcal(p)}{\int_0^1 \Pcal(\text{data}|p)\,\Pcal(p)\,\dif p}
\ee
or rather
\begin{align*}
\Pcal_{\text{post,1}}(\qnoinf|k_1)
&= \frac{\Pcal(k_1|\qnoinf)\ \Pcal_{\text{prior}}(\qnoinf)}{\int_0^1 \Pcal(k_1|\qnoinf)\,\Pcal_{\text{prior}}(\qnoinf)\,\dif \qnoinf} \\
&= \frac{\left[(1-\qnoinf^{\Dil_1})^{k_1}\, \qnoinf^{\Dil_1(n_1-k_1)}\right]\, \Pcal_{\text{prior}}(\qnoinf)}{\int_0^1 \Pcal(k_1|\qnoinf)\,\Pcal(\qnoinf)\,\dif \qnoinf} \\
\Pcal_{\text{post,1}}(\qnoinf|k_1)
&\propto \left[(1-\qnoinf^{\Dil_1})^{k_1}\, \qnoinf^{\Dil_1(n_1-k_1)}\right]\, \Pcal_{\text{prior}}(\qnoinf) 
\end{align*}
where $\Pcal_{\text{post,1}}(\qnoinf|k_1)$ is our updated, posterior belief about $\qnoinf$ after having observed $k_1$ successes out of $n_1$ trials in the first column ($i=1$), and given our prior belief, $\Pcal_{\text{prior}}(\qnoinf)$, about $\qnoinf$ before making this observation.

\subsubsection{Considering all dilutions of the \ID assay}

As mentioned above, in the 96-well \ID assay, each dilution contains a number of independent infection wells (replicates) inoculated with the same sample concentration. This process is then repeated over a series of dilutions, each separated from the previous by a fixed dilution factor. Having observed the fraction of wells infected at the first dilution considered, $\Dil_1$, we have updated our posterior belief about \qnoinf. We will now use this updated belief as our new prior as we observe our second dilution ($\Dil_2$), such that
\begin{align*}
\Pcal_{\text{post,2}}(\qnoinf|\vec{k}_2)
&\propto \Pcal(k_2|\qnoinf)\ \Pcal_{\text{post,1}}(\qnoinf|k_1) \\
\Pcal_{\text{post,2}}(\qnoinf|\vec{k}_2)
&\propto \left[(1-\qnoinf^{D_2})^{k_2}\, \qnoinf^{D_2(n_2-k_2)}\right]\, \left[(1-\qnoinf^{\Dil_1})^{k_1}\, \qnoinf^{\Dil_1(n_1-k_1)}\right] \Ppri(\qnoinf) \\
\Pcal_{\text{post,2}}(\qnoinf|\vec{k}_2)
&\propto \Qcal(\vec{k}_2|\qnoinf)\, \Ppri(\qnoinf)\ ,
\end{align*}
where we introduce $\vec{k}_2 = \{k_1,k_2\}$ and
\be
\Qcal(\vec{k}_2|\qnoinf) = \left[(1-\qnoinf^{D_2})^{k_2}\, \qnoinf^{D_2(n_2-k_2)}\right]\, \left[(1-\qnoinf^{\Dil_1})^{k_1}\, \qnoinf^{\Dil_1(n_1-k_1)}\right]
\ee
as short-hands for convenience. From this, it is easy to extrapolate the posterior likelihood distribution (pPLD) after having observed all $J$ dilutions ($\Dil_1$, $\Dil_2$, ..., $\Dil_J$) of the \ID assay, namely
\begin{equation}
\Pcal_{\text{post,J}}(\qnoinf|\vec{k}_{J})
\propto \Qcal(\vec{k}_J|\qnoinf)\, \Ppri(\qnoinf)
\label{eqn:ppostq}
\end{equation}
where
\begin{equation}
\Qcal(\vec{k}_J|\qnoinf) = \left[\prod_{j=1}^{J} (1-\qnoinf^{\Dil_j})^{k_j}\right] \qnoinf^{\sum_{j=1}^{J} \Dil_j(n_j-k_j)} \ .
\label{eqn:q11}
\end{equation}
Note that this expression is largely equivalent to that obtained by Mistry et al.\ \cite{mistry18}.

\subsubsection{Considering the choice of prior}

In Eqn.\ \eqref{eqn:ppostq}, we obtained a pPLD for \qnoinf. Our objective, however, is to estimate the pPLD of $\Cvir$, the specific infection concentration in our sample, rather than $\qnoinf$. In fact, because both the plaque and \ID assays provide an accuracy that is normally distributed in $\log_{10}(\Cvir)$ rather than $\Cvir$, it follows that $\log_{10}(\Cvir)$ (hereafter \lCvir) rather than \Cvir is the quantity of interest. We note that $\Qcal(\vec{k}_J|\qnoinf)$ in Eqn.\ \eqref{eqn:ppostq} is a probability density function in $\vec{k}_J$ rather than in $\qnoinf$. As such, a change of variables, say from $\qnoinf$ to $\lCvir(\qnoinf)$, would affect only the prior because $\Qcal(\vec{k}_J|\qnoinf) = \Qcal(\vec{k}_J|\qnoinf(\lCvir)) = \Qcal(\vec{k}_J|\lCvir)$. Thus, the pPLD for $\lCvir$ is given by
\begin{equation}
\Pcal_{\text{post},J}(\lCvir|\vec{k}_J) \propto \Qcal(\vec{k}_J|\qnoinf(\lCvir))\ \Ppri(\lCvir)\ ,
\label{eqn:lcvpri}
\end{equation}
where $\Qcal(\vec{k}_J|\qnoinf)=\Qcal(\vec{k}_J|\Cvir)=\Qcal(\vec{k}_J|\lCvir)$ because $\Qcal(\vec{k}_J|\qnoinf(...))$ can be written in terms of $\qnoinf$, $\Cvir$, or $\lCvir$, because it is a probability density function in $\vec{k}_J = \{k_1,k_2,...,k_{J}\}$ rather than in $\qnoinf$. To complete this expression, we need to choose a physically and biologically appropriate prior belief regarding \lCvir.

Prior to conducting the \ID assay, we know at least that $\Cvir\in[1/V_\text{Earth},1/\Vvir]$, where $1/\Vvir$ is the maximum possible concentration, namely that if the entire volume of the sample is constituted solely of infectious virions, and $1/V_\text{Earth}$ is the minimum possible concentration, namely that if there was only one infectious virion left on Earth. As we explain below, these limits are not important; only the fact that they are convincingly physically bounded both from above and below, i.e., $\in(0,\infty)$, is relevant.

If we choose our prior to be uniform in $\Cvir\in[1/V_\text{Earth},1/\Vvir]$, namely $\Ppri(\Cvir)=1/(1/\Vvir-1/V_\text{Earth})\approx\Vvir$, and using the fact that $\Ppri(\Cvir)\,\dif\Cvir = \Ppri(\lCvir)\,\dif\lCvir$, we can write
\be
\Ppri(\lCvir) = \Ppri(\Cvir) \frac{\dif\Cvir}{\dif\lCvir} = \Vvir \frac{\dif\left[10^{\lCvir}\right]}{\dif\lCvir} = \Vvir \ln(10) 10^{\lCvir} \propto 10^{\lCvir}
\ee
which yields
\begin{equation}
\Pcal_{\text{post},J}(\lCvir|\vec{k}_J) \propto \Qcal(\vec{k}_J|\qnoinf(\lCvir))\ 10^{\lCvir}\ .
\label{eqn:postCvir}
\end{equation}
We see here that the range chosen for the uniform prior in $\Cvir$ is not important because it only contributes a constant to our proportionality Eqn.\ \eqref{eqn:lcvpri}.

Alternatively, because the \ID assay estimates $\lCvir$ rather than $\Cvir$, our prior belief about the virus concentration is more appropriately expressed in $\lCvir$ rather than $\Cvir$. Again, the bounds of the uniform distribution in \lCvir is unimportant, provided that it is finite in extent such that $\lCvir\in[{\lCvir}_\text{min},\log_{10}(1/\Vvir)]$ where ${\lCvir}_\text{min}>-\infty$, such that we can write
\begin{equation}
\Pcal_{\text{post},J}(\lCvir|\vec{k}_J) \propto \Qcal(\vec{k}_J|\qnoinf(\lCvir))\ .
\label{eqn:postlC}
\end{equation}

\begin{figure}
\includegraphics[width=\textwidth]{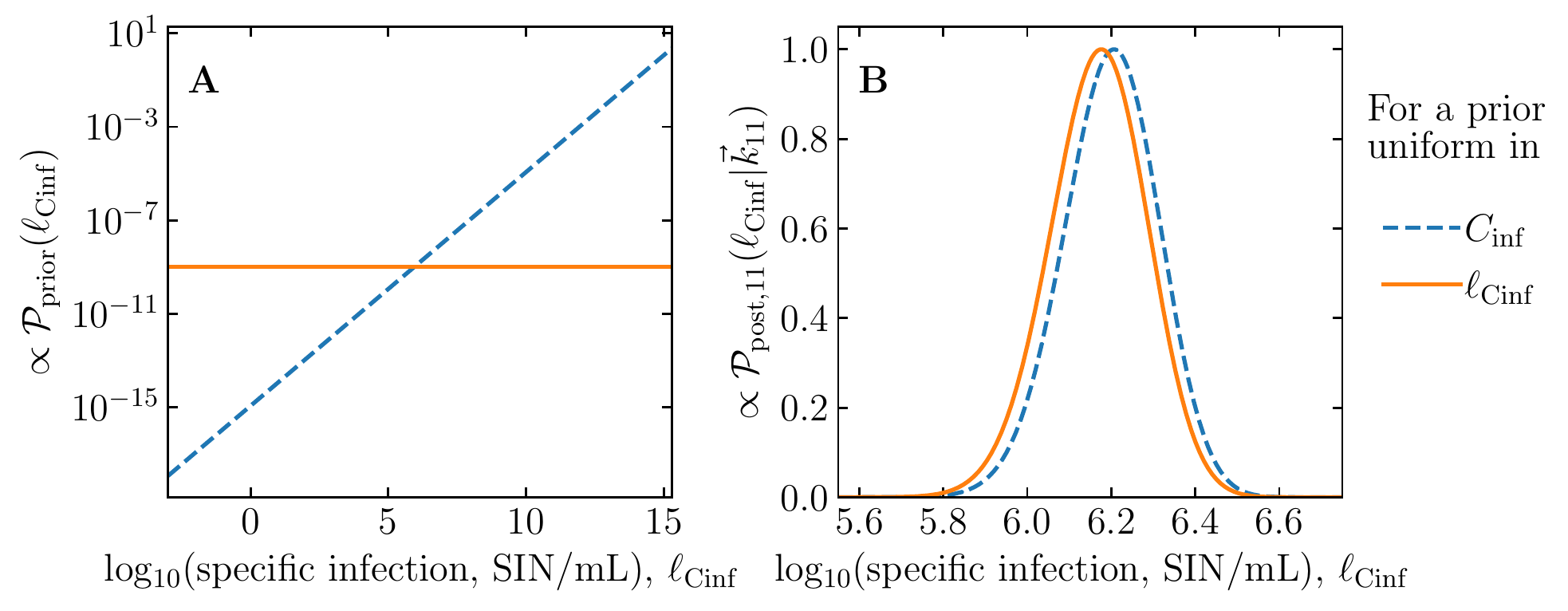}
\caption{%
\textbf{Impact of the choice of prior on the posterior distribution for $\lCvir$.} %
(A) Non-normalized priors for $\log_{10}$(specific infections, \idnew/mL)$=\lCvir$ that are uniform in either \Cvir or \lCvir are shown. A prior uniform in \Cvir is biased towards larger values of \lCvir. %
(B) Updated posterior belief about \lCvir for each of the two prior beliefs shown in (A), as per Eqns.\ \eqref{eqn:postCvir} and \eqref{eqn:postlC}, after having observed the \ID assay example provided in Section \ref{resone}. While the prior uniform in \Cvir yields a pPLD with a mode of $\lCvir=6.21$, that for a prior uniform in \lCvir yields a mode of $\lCvir=6.18$.
}
\label{fprior}
\end{figure}

Figure \ref{fprior} illustrates the two distinct priors assumed to arrive at Eqns.\ \eqref{eqn:postCvir} and \eqref{eqn:postlC} and their impact on the posterior $\Pcal_{\text{post},J}(\lCvir|\vec{k}_J)$ for the example \ID experiment described in Section \ref{resone}. Figure \ref{fprior}A illustrates the consequence of choosing a prior uniform in $\Cvir$, i.e., a bias towards higher virus concentrations. This is because a uniform prior in \Cvir corresponds to a belief that one is as likely to measure a set of virus concentrations in the range $[0.001,\, 0.002]$ as in the range $[1,000,000.001,\, 1,000,000.002]$. When plotted on a log-scale, there are $100\times$ more intervals of width 0.001 in $[10^4,10^5]$ than in $[10^2,10^3]$. Thus, this prior corresponds to a belief that the likelihood of measuring a certain virus concentration increases exponentially as \lCvir increases linearly. In contrast, a prior uniform in \lCvir corresponds to a belief that one is as likely to measure a set of virus concentrations in the range $[0.001,\, 0.002]$ than in the range $[1,000,000,\, 2,000,000]$, or rather in the range $[1,2]\times10^{-3}$ than in the range $[1,2]\times10^{6}$. As such, a uniform distribution in \lCvir is more physically and biologically sensible and therefore was chosen for our estimation method.

\subsection{Calculation of \newprog's outputs}

One of the graphical outputs of \newprog is the non-normalized PLD of \lCvir given the number of wells that were infected at each dilution, $\vec{k}_{J}$, like that shown in Figure \ref{fig:example1}(left panel), computed as
\begin{align}
\Upost(\,\lCvir\,|\,\vec{k}_{J}\,) &= \prod_{j=1}^{J}\ 
 \frac{n_j!}{k_j!\ (n_j-k_j)!}\ \cdot\ p_j^{k_j}\ \cdot\ (1-p_j)^{n_j - k_j} \\
\text{where}\ \ 
p_j &= 1 - \exp\left[\ -10^{\lCvir} \cdot \Vinoc \cdot \Dil_j\ \right] \ .
\end{align}
While \Upost is not the normalized likelihood of \lCvir, its maximum value at its mode (${\lCvir}_\text{,mode}$) is the normalized probability of observing this particular \ID plate outcome ($\vec{k}_{J}$) out of all other possible plate outcomes, assuming the true, specific infection concentration in the sample is ${\lCvir}_\text{,mode}$.

Another visual output of \newprog is a graphical representation of the theoretical number of wells that would be infected given the most likely \lCvir, like that shown in Figure \ref{fig:example1}(right panel). It is computed following
\begin{equation}
N_\text{wells infected}(x) = N_\text{wells total} \left[1 - \exp\left(-10^{{\lCvir}_\text{,mode}}\, \Vinoc\, 10^{-x}\right)\right] \ , \label{eqn:Ntheo}
\end{equation}
where $x$ is the $\log_{10}$ of the dilution such that $\Dil = 10^{-x}$ is the dilution. It corresponds to the continuous equivalent of this quantity which is discrete in the \ID assay, namely $\Dil_i=10^{-x_i}$ which is the $i^\mathrm{th}$ dilution of the sample. As such, $\Dil_i =$ (minimum dilution) $\cdot$ $\text{(dilution factor between columns)}^{i-1}$ where $i\in{[1,J]}$. For example, if the dilution of the least diluted column is $0.1=10^{-1}$ and the dilution factor between dilutions in the \ID assay is such that it halves the concentration between each dilution, i.e., $1/2 = 2^{-1} = 10^{-\log_{10}(2)} \approx 10^{-0.301}$, then $\Dil_i = 10^{-1}\cdot10^{-0.301\cdot(i-1)}$ such that $\Dil_1 = 10^{-1}$, $\Dil_2 = 10^{-1.301}$, $\Dil_3 = 10^{-1.602}$, and so on, such that $x_1 = 1$, $x_2 = 1.301$, $x_3 = 1.602$, and so on.

In the graphical representation of the \ID assay, the edges of the grey bands flanking the theoretical blue curve correspond to Eqn.\ \eqref{eqn:Ntheo} wherein ${\lCvir}_\text{,mode}$ has been replace by the 68\% and 95\% CI values for $\lCvir$. These CI bands \emph{do not} correspond to the 68\% and 95\% CI of the expected number of infected wells at each dilution given ${\lCvir}_\text{,mode}$.

The sample dilution corresponding to \unit{1}{\tcid} estimated based on the biased \RM and \SK approximations (right panels) are converted to \idnew\ (left panels) based on $\unit{1}{\tcid} = \unit{\me{\gamma=0.5772}}{\idnew} = \unit{1.781}{\idnew}$ \cite{wulff12,govindarajulu01}. In contrast, the \lidnew computed by \newprog can be converted to a true (unbiased) estimate of $\log_{10}(\tcid)$ using $\unit{1}{\tcid} = \unit{1/\ln(2)}{\idnew} = \unit{1.44}{\idnew}$ \cite{bryan1957}.

\subsection{Infection concentration measures of influenza A virus samples}

\subsubsection{Cell culture}

Madin-Darby canine kidney cells (MDCKs) were cultured in growth media (complete MEM media with 5\% heat-inactivated FBS), in tissue culture treated T75 flasks, at 37\degree C with 5\% CO$_2$ and 95\% relative humidity. Cells were split 1/10 every 3--4 days or upon reaching approximately 95\% confluency. One passage of cells was expanded for use by both researchers in one experiment to quantify the 50\% tissue culture infectious dose (\tcid) and plaque forming units (\pfu) of one viral strain.

\subsubsection{Viral stocks}

Stocks of influenza A/Puerto Rico/8/34 (H1N1) (PR8) and influenza A/California/4/09 (Cali/09) were stored at -80\degree C and thawed on ice immediately before use. The \tcid\ and \pfu\ of stock viruses was known to both researchers prior to this study. Serial dilutions were made in MDCK infection media (complete MEM media with 4.25\% BSA) and dilutions were made by each researcher independently for titering. `Researcher A' and `Researcher B' independently performed the \tcid\ and \pfu\ assays of one viral strain for one experiment on the same day using the same viral stock, reagents, and passage of cells. Each experiment was performed on a separate day (Fig.\ \ref{fig:amber-pfu}).

\subsubsection{Plaque assay}

MDCKs were seeded in six-well plates (\unit{5.5\times10^5}{cells/\milli\liter}, \unit{2}{\milli\liter/well}) and grown to 90\% confluency overnight (37\degree C, 5\% CO$_2$, 95\% relative humidity). Each six-well plate contained 10-fold serial dilutions plated in singlet as well as a negative control and five 6-well plates were carried out per experiment. Cells were washed twice with PBS w/ Ca$^{2+}$Mg$^{2+}$ before the addition of \unit{500}{\micro\liter} of viral dilutions per well. After \unit{1}{\hour} at room temperature on a rocker, the inoculum was aspirated, cells were washed with PBS containing Ca$^{2+}$Mg$^{2+}$ (PBS w/ Ca$^{2+}$Mg$^{2+}$) (Gibco), and gently covered with \unit{2}{\milli\liter} of agarose overlay (complete media, 4.25\% BSA, 0.9\% agarose, \unit{1}{\micro\gram/\milli\liter} TPCK-Trypsin). After drying the overlay at room temperature, plates were inverted and incubated (37\degree C, 5\% CO$_2$, 95\% relative humidity) for \unit{3}{d} (PR8) or \unit{4}{d} (Cali/09). Plaques were visualized by staining cells with 0.1\% crystal violet solution in 37\% formaldehyde for \unit{30}{min} and counted by `Researcher A' or `Researcher B' on their respective experiments (Fig.\ \ref{fig:amber-pfu}).

\subsubsection{\tcid\ assay}

MDCKs were seeded in 96-well flat bottom plates ($5\times10^4$\,cells/\unit{100}{\micro\liter}, \unit{100}{\micro\liter/well}) and grown to 80\% confluency overnight (37\degree C, 5\% CO$_2$, 95\% relative humidity). For each experiment, 4 replicate wells, at each of 7 different dilutions separated by a 10-fold dilution, were infected, and the dilution series was performed 5 times. Cells were washed with PBS w/ Ca$^{2+}$Mg$^{2+}$ before the addition of \unit{100}{\micro\liter} of viral dilutions per well. After \unit{1}{\hour} at room temperature on a rocker, the inoculum was aspirated and replaced with \unit{100}{\micro\liter} of infection media containing \unit{1}{\micro\gram/\milli\liter} TPCK-Trypsin. Cells were incubated (37\degree C, 5\% CO$_2$, 95\% relative humidity) for \unit{3}{d} (PR8) or \unit{4}{d} (Cali/09). Supernatants were used to do a hemagglutination (HA) assay with chicken red blood cells. HA assays were performed and read by `Researcher A' or `Researcher B' on their respective experiments.

\subsubsection{Statistical analysis}

The data points reported in Figure \ref{fig:amber-pfu}C,D were computed by taking each of the 5 replicates measured with either the PFU, RM, or SK and the 5 replicates measured via \idnew\ (5 replicates $\times$ 5 replicates = 25 pairs) for each of the 2 experiments by each of the 2 researchers, yielding 100 pairs. For each pair, the $\log_{10}$ of ratio of either PFU, RM or SK over SIN was computed. The mean and standard deviation of the resulting 100 $\log_{10}(\mathrm{ratio})$ were computed and are reported in Figure \ref{fig:amber-pfu}C,D. The statistical significance ($p$-value) of the differences between (PFU,RM,SK) and (\idnew) was computed using the Mann-Whitney U test (\texttt{scipy.stats.mannwhitneyu}).


\cleardoublepage
\bibliographystyle{abbrvurl}
\bibliography{measure-sin}


\end{document}